\documentclass[twocolumn,fleqn]{article}

\begin{document}
\twocolumn[
{\Large\bf
Relation between confinement and higher symmetry
restrictions \\ for color  particle motion     
}

\bigskip

{\large V.V. Khruschov}

\medskip

{\it Centre for Gravitation and Fundamental
Metrology, VNIIMS, 46 Ozyornaya St.,
Moscow 119361, Russia
}

\medskip

\vspace*{-0.5cm}

\begin{center}
\begin{abstract}
\parbox{15cm}{Quantum operators of coordinates and  momentum components
 of a particle in  the Minkowski spacetime can belong to the
generalized Snyder-Yang algebra  and produce a quantum phase space 
with three new constants in the general case. With account for the O(2,6)
invariance in the quantum phase space of a color particle
the equation of motion is obtained, which contains an oscillator
rising potential. The presence of the oscillator
 potential can simulate a confinement of a color particle.
A parameter of the oscillator potential is estimated and a relationship
between current and constituent quark masses is obtained.

\medskip
\noindent{\it Keywords}: spacetime symmetry; quantum constant;
 quantum phase space; quark; confinement

\smallskip
\noindent{PACS:} 11.30.Ly; 11.90.+t; 12.90.+b

}

\end{abstract}
\end{center}
\bigskip
\smallskip
]

\smallskip

\noindent {\large\bf 1. Introduction}

\smallskip

 Now it is  generally accepted that QCD is the theory of strong 
interaction of quarks and gluons. As is known  QCD operates with 
 quantum color fields of quarks and gluons defined in the conventional
four dimensional Minkowski spacetime $M_{1,3}$ \cite{bog,yndu}.  QCD has 
considerable verification at high interaction energies,
however some problems remain unsolved in the low energy region, 
for instance, a confinement of color particles and a violation 
of chiral invariance of the massless QCD lagrangian.

It is well known that now the quark confinement is investigated in the frame 
of different approaches such as lattice QCD, Schwinger-Dyson equations, 
massive transverse gluons, potential models, {\it etc} 
\cite{green,swan1,swan2,alk}. In the present letter we consider a model for 
resolving the confinement problem with the help of higher symmetry restrictions
for a  motion of a color particle.  To do this would
require an additional prerequisite, namely,  the O(2,6)
invariance in a phase space of a color particle, which have been
proposed in ref.\cite{khru1}. Besides we use the quantum phase space
with three fundamental constants additional to the standard ones $c$ 
and $\hbar$ \cite{sny,yang,lezkh}.

\smallskip

\noindent {\large\bf 2. Restrictions for color particle motion based on higher 
symmetry in a quantum phase space}
 
\smallskip

Since coordinates and momentum components of a quantum particle
can be noncommutative in the general case, let us start with the
generalized Snyder-Yang algebra (GSYA) to be 
considered in the following  form \cite{sny,yang,lezkh,khru2}:

\[
[F_{ij}, F_{kl}]=i(g_{jk}F_{il}-g_{ik}F_{jl}+g_{il}F_{jk}-g_{jl}F_{ik}),
\]
\[
[F_{ij}, p_{k}]=i(g_{jk}p_{i} - g_{ik}p_j), 
\]
\[ [F_{ij}, q_k]=i(g_{jk}q_i - g_{ik}q_j),
\]
\begin{equation}
[F_{ij}, I]=0, \quad [p_i, q_j]=i(g_{ij}I + \kappa F_{ij}),
\label{al1}
\end{equation}
\[  
[p_i, I]=i(\mu^2q_i - \kappa p_i),   [q_i, I]=i(\kappa q_i-\lambda^2p_i),
\]
\[
[p_i, p_j ]=i\mu^2F_{ij},  \quad [q_i, q_j]=i\lambda^2F_{ij}, 
\]

\noindent where $c = \hbar = 1$,  $F_{ij}$, $p_i$, 
$x_i$ are the generators of the Lorentz group and the operators of momentum 
components and coordinates, correspondingly, $I$ is the "identity" operator, 
$i, j, k, l = 0, 1, 2, 3$.  The new quantum constants $\mu$
and $\lambda$ have dimensionality of mass and lenght correspondingly.
The constant $\kappa$ is dimensionless in the natural system of units.

By applying the algebra (\ref{al1}) to the description of color particles
the condition $\kappa = 0$ can be imposed. Actually it is known 
the  nonzero $\kappa$ leads to the $CP-$violation \cite{lezkh}, 
but  strong interactions   are invariant 
with respect to the $P-$, $C-$ and $T-$transformations on the high level 
of precision.  Moreover for color particles one can use the relation
$\mu\lambda = 1$\footnote{In Ref.\cite{khru2} the relation $ \lambda = 0$ 
is used  instead of $\mu\lambda = 1$.}. In this case  we obtain the 
reduction of GSYA to the special Snyder-Yang algebra (SSYA) with 
$\mu\lambda = 1$ and $ \kappa = 0$  for strong interaction color particles. 
 Denoting $\mu$  as $\mu_c$ and  $\lambda$ as $\lambda_c$ we write
 the following  commutation relations without the standard commutation
relations with Lorentz group generators, which are  shown 
 for the GSYA above (see eqs.(\ref{al1})).

\[
[p_i, q_j]=ig_{ij}I, [p_i, I]=i\mu_c^2q_i,  [q_i, I]=-i\lambda_c^2p_i,
\]
\begin{equation}
  [q_i, q_j]=i\lambda_c^2F_{ij},  \quad [p_i, p_j ]=i\mu_c^2F_{ij}.
\label{al2}
\end{equation}

We take into account difficulties arised when one try to prove 
the confinement on the basis of the QCD first principles, so we 
simulate  this phemomenon with the help of an
  assumed high symmetry of the nonperturbative OCD interaction.
We suppose that  the nonperturbative OCD interaction have the property
of an approximated or exact  higher spacetime symmetry beyond the
 Poincare symmetry.
In our model  we turn from the  Poincare symmetry in the
 Minkowski spacetime  to the inhomogeneous O(2,6) symmetry in a 
phase space of a color particle \cite{khru1}.

    In this way we consider the generalized  model for a color particle motion,
when coordinates and momenta are on equal terms and form an eight dimensional  
 phase space: $h=$$\{h^{A}|h^{A}=q^{\mu},A=1,2,3,4,$ $ \mu =0,1,2,3,$
$h^{A}=$ $\tau p^{\mu},A=5,6,7,8,$ $ \mu =0,1,2,3\}$. 
$P=$ $\{P^{A}|P^{A}=$ $p^{\mu},A=1,2,3,4,$ 
$ \mu =$ $0,1,2,3,$ $P^{A}=$ $\sigma q^{\mu},A=$ $5,6,7,8,$ $ \mu =0,1,2,3\}$.
The constants $\tau$ and $\sigma$ have dimensions of length and 
mass square, correspondingly. Their values can be chosen on the 
phenomenological ground or with the help of some functions of
the quantum constants  $\mu$, $\kappa$ and $\lambda$.
So the generalized lenght square 
\begin{equation}
  L^2=h^Ah_A,
\end{equation}
\noindent and the generalized mass square 
\begin{equation}
  M^2=P^AP_A,
\end{equation}
\noindent are invariant under the O(2,6) transformations, 
where $h_A=g_{AB}h^B$,
$g_{AB}=$ $ g^{AB}=$ $diag\{1,-1,-1,-1,1,-1,-1,-1\}$.

Thus we propose that for strong interacting color particles 
the  generalized differential mass squared  has the physical meaning:
\[
dM^2 = (dp_0)^2 -(dp_1)^2 - (dp_2)^2 - (dp_3)^2
\]
\[
+ \sigma^2(dq_0)^2-\sigma^2(dq_1)^2 -\sigma^2(dq_2)^2 -\sigma^2(dq_3)^2 
\]
\begin{equation}
= (dm)^2+\sigma^2(ds)^2.
\end{equation}
\noindent 
An important point is that the coordinates $q^{\mu}$ and the momentum  
components $p^{\mu}$ are the quantum operators satisfied eqs.(\ref{al1})
 or eqs.(\ref{al2}) in the frame of this approach.

Under these conditions the new Dirac type equation for a spinorial
field $\psi$ has the following form:
\begin{equation}
  \gamma^AP_A\psi = M\psi,
\label{newd}
\end{equation}
\noindent where 
$\gamma^A$ are the Clifford numbers for the 
spinorial O(2,6) representation, i.e.
\begin{equation}
  \gamma^A\gamma^B+\gamma^B\gamma^A = 2g^{AB}.
\end{equation}

One can take the product of eq.(\ref{newd}) with 
$\gamma^AP_A+M$ and apply eqs.(\ref{al1}), then the following 
equation for $\psi$ can be obtained
\[
(p^ip_i + \sigma^2q^iq_i + 2\Sigma_{i<j}S^{ij}F_{ij} +
\]
\begin{equation}
  + 2\sigma S^0I)\psi = M^2\psi,  S^{0}=\frac{i}{2}C^{0}, 
S^{ij}=\frac{i}{2}C^{ij},
\label{newds}
\end{equation}
\noindent where
\[
C^0 = \gamma^1\gamma^5g_{00}+\gamma^2\gamma^6g_{11} 
+ \gamma^3\gamma^7g_{22}+\gamma^4\gamma^8g_{33},
\]
\[
C^{01} =\gamma^1\gamma^2\mu^2+\gamma^5\gamma^6\sigma^2\lambda^2
+ (\gamma^1\gamma^6-\gamma^2\gamma^5)\sigma\kappa,
\]
\[
C^{02} =\gamma^1\gamma^3\mu^2+\gamma^5\gamma^7\sigma^2\lambda^2
+ (\gamma^1\gamma^7-\gamma^3\gamma^5)\sigma\kappa,
\]
\[
C^{03} =\gamma^1\gamma^4\mu^2+\gamma^5\gamma^8\sigma^2\lambda^2
+ (\gamma^1\gamma^8-\gamma^4\gamma^5)\sigma\kappa,
\]
\[
C^{12} =\gamma^2\gamma^3\mu^2+\gamma^6\gamma^7\sigma^2\lambda^2
+ (\gamma^2\gamma^7-\gamma^3\gamma^6)\sigma\kappa,
\]
\[
C^{13} =\gamma^2\gamma^4\mu^2+\gamma^6\gamma^8\sigma^2\lambda^2
+ (\gamma^2\gamma^8-\gamma^4\gamma^6)\sigma\kappa,
\]
\begin{equation}
 C^{23} =\gamma^3\gamma^4\mu^2+\gamma^7\gamma^8\sigma^2\lambda^2
+ (\gamma^3\gamma^8-\gamma^4\gamma^7)\sigma\kappa.
\label{ewds}
\end{equation}

Eq.(\ref{newds}) contains the oscillator potential, which restricts
a motion of a color quark. Besides that we broke the inhomogeneous
 O(2,6) symmetry with the help of the commutation relations (\ref{al1}).
In the special case $\mu\lambda = 1$, $ \kappa = 0$ and the commutation
 relations (\ref{al2}) for SSYA we will obtain more 
simple expressions for the $C^0$ and $C^{ij}$, but the form of the  
eq.(\ref{newds}) will remain unchanged. Note that  eq.(\ref{newds}) can also be
applied for a description of a confinement of boson particles such as 
diquarks and gluons with the same confinement parameter $\sigma$.

Let us consider some consequences of this approach for 
specific  color quark characteristics.
   From the relations (\ref{al2}) it immediately follows nonzero
uncertainties for results of simultaneous measurements of quark
momentum components. For instance, let $\psi_{1/2}$ is 
a quark state with a definite value of its spin component along the third axis.
Consequently, $[p_1, p_2] = i\mu_c^2/2$, thus
\begin{equation}
\Delta p_1 \Delta p_2 \ge \mu_c^2/4
\end{equation}
     and if $\Delta p_1 \sim \Delta p_2$, one gets $\Delta p_1 >\mu_c/2$,
$\Delta p_2 >\mu_c/2$. 
 We see that the generalized quark momentum components $p_{\perp1,2}$ cannot be 
measured better than tentatively one-half a  value of $\mu_c$ \cite{khru1}.

One can get an  estimation of the $\sigma$ value 
  using the quark equation (\ref{newds}). As it is seen,  
$M^2$ and $p^2$ entered into the eq.(\ref{newds})
can be considered as current and constituent  quark masses squared, 
respectively.  So eq.(\ref{newds}) indicates that the convential 
relation $p_{cur}^2=M^2$ for a current quark should be transform
to $p^2=M^2+ \Delta^2$ for a constituent quark, where $m^2=M^2+\Delta^2$
is a constituent mass squared.
To estimate the $\sigma$ value with the help of 
a constituent quark mass $m$ and a current quark mass $M$  values 
 a ground state $\psi_0$  in a meson has been considered  neglecting
 the orbital angular momentum contribution $L\psi_0$. 
By this means in ref.\cite{khru2}, where another equation for 
a generalized quark has been used, it has been obtained that 
the $\mu_c$ (or $\sigma$) value is of the order of 0.2 $GeV$. 

 Let us estimate the $\sigma$ value
with the help of the value of the  confinement rising  potential coefficient 
\cite{khru3}.
One can see from eq.(\ref{newds}) that the coefficient of the oscillator 
confinement  potential for a color particle is equal to $\sigma^2$. 
In the  one particle potential approach this coefficient is connected 
with the so-called
 "string tension" $\sigma_{str}$  typically as $\sigma^2=\sigma_{str}^2/4$, 
where $\sigma_{str}$ varies from 0.19 $GeV^2$
to 0.21 $GeV^2$ \cite{khru3}. Hence $\sqrt{\sigma}$ $\approx 0.3 GeV $  and 
$\lambda_{conf}\approx 0.6 Fm$. Clearly  it is assumed that the 
conceptions of the asymptotical  oscillator potential and the constituent quark
are applicable in a confinement domain. 

\smallskip
\noindent {\large\bf 3. Conclusion}
\smallskip

Above we considered the model of a color particle confinement
with  the generalized O(2,6) symmetry in the quantum phase space 
of a color particle. In the framework of the model the new equations
of motion  (\ref{newd}) and (\ref{newds}) for  color particles
have been obtained with the oscillator rising potential which provides 
the confinement of the particles.
 The  further investigation of properties  of  solutions  of   the
equations  (\ref{newd}) and (\ref{newds})  is the important objective
 of this model and under consideration now.

\end{document}